# GAMMA RADIATION SPECTRA OF 1200 MeV ELECTRONS IN THICK BERYLLIUM, SILICON AND TUNGSTEN SINGLE CRYSTALS

*G.L.Bochek, O.S. Deiev, V.L. Kulibaba, N.I. Maslov, V.D. Ovchinnik, B.I. Schramenko*

*National Science Center "Kharkov Institute of Physics and Technology", 61108, Kharkov, Ukraine*

Gamma radiation spectra of 1200 MeV electrons in the single crystals of the beryllium 1.2 mm thick, silicon 1.5 mm and 15.0 mm thick and tungsten 1.18 mm thick along of the crystallographic axes were measured. Also spectral-angular distributions of gamma radiation in the silicon single crystals 1.5 mm thick along of the crystallographic axes <100>, <110> and <111> were measured. On the basis of these measurements the radiation spectra for the different solid angles up to $6.97 \times 10^{-6}$ sr were obtained. Gamma radiation spectra of the electrons with energy 1200, 600 and 300 MeV in the Si single crystals 15.0 mm thick and the W thick 3.0 mm also were measured. The above results complement the previously obtained experimental and theoretical data and can be used to create a source of intense gamma-ray beams for radiation technologies.

## 1. INTRODUCTION

In recent decades the electromagnetic radiation (x-ray and gamma range) of the relativistic electrons in single crystals is object of numerous theoretical and experimental studies for the electrons energy from several MeV to hundreds GeV.

Interest to the channeling radiation of the electrons was amplified after the idea about a possibility of its application as intensive and easily tunable source quasimonoenergetic X-ray and gamma radiation [1-15] and also as the positrons source for the linear supercolliders [16].

For example, at the plane channeling of electrons with energy about 50 MeV in the radiation spectra intensive narrow spectral lines are observed. They can find the use in applications which require the intensive and quasimonoenergetic the gamma radiation source.

In the paper [2] the detailed theoretical analysis was performed of all possible factors wich influence on the parameters of an intensive X-ray source (atomic number, crystal structure, heat conductivity, radiation resistance etc.). Such source of radiation can based on the channeling radiation in Be, C, Si, Ge and W single crystals and the high-current electron beams with energy about 50 MeV.

In particular, it is shown that from all considered materials the diamond crystal provides of the greatest gamma radiation intensity. However, Be crystal not much more concedes to a diamond crystal.

At energy range of electrons 10...110 MeV systematic experimental study of gamma radiation spectra at the electrons plane channeling was performed for Si, C and Be crystal [12]. The main attention is paid to optimization of the yield and the spectral line width of the gamma radiation [2,11, 13-15].

Range of the electrons energy of about 1000 MeV is interesting by that in the gamma radiation spectra of electrons at their moving along crystallographic axes is observed noticeable increasing of the yield of gamma quanta with energy in the region of "giant resonance" in the cross section of photonuclear reactions.

Transition from traditional amorphous converters to the crystal targets of optimal thickness can help to solve the problem of the transformation coefficient increase of charged particles (electrons) energy in the gamma radiation energy [17-19].

It is known that, at initial electrons energy 1200 MeV for the silicon crystal 15.0 mm thick the transformation coefficient reaches 6% that significantly exceeds the value of this parameter for the amorphous target This circumstance can be used in various applied and scientific nuclear and physical researches [17, 20-25].

The comprehensive theoretical analysis of the experimental data obtained on linacs in various scientific centers (KIPT, SLAC, YerPhI, NPI TPU etc.) was performed. This made it possible to conclude about the considerable (in case of electrons – crucial) influence of the dynamics of the charged particles beams in the thin crystal on the formation of the main gamma radiation characteristics, such as the spectrum and intensity.

In thick crystals as a result of the process of electrons multiple scattering and their fast dechanelling [26], the main radiation mechanism is the emission of overbarrier electrons. The important role of particle beam dynamics in a crystal in determining the main radiation mechanisms was clarified, and, as a consequence, the main characteristics of this radiation were evaluated.

Presently the new opportunities for obtaining the intensive electromagnetic radiation of the electrons when using bent crystals and crystal undulators are studied. The radiation of electrons was studied experimentally for the electrons energy 855 MeV [27], and theoretically by means of simulation [28].

It should be noted that all the experimental results known so far were obtained at very small electrons currents. In this conditions no more than one photon in an accelerator current impulse are registered.

However, during creation of the intensive gamma radiation sources there can be difficulties because of big heat load on the single crystals when using high-current electron beams.

The investigation of the influence of the impulse current of the linac beam on the gamma radiation yield from the single crystal at the electrons channeling and calculation of the ionization energy losses of the electrons with energy 300 and 1200 MeV in the silicon single crystal 15 mm thick was given in paper [29].

It is shown that in the range of impulse current 1…100 mA the gamma radiation yield of the 1200 MeV electrons increases in proportion to the value of the impulse current. This demonstrates the preservation by the crystal converter of the single crystal properties up to the impulse current of electrons 100 mA.

Similar investigations at the low electrons energy are given in papers [5-8] and in the recent years in papers [13-15].

For various tasks it is important to have the gamma quanta beams with small divergence and the maximal brightness in the given spectral interval. For such problems it is necessary to know the spectral and angular distribution of the gamma radiation and find optimum parameters of the beam - crystal system on the corresponding exit parameters.

The purpose of the given work is further studying of the gamma radiation characteristics of the electrons from single crystals for the solution of such problems as the choice of the crystal material, the crystallographic axes and minimum energy of electrons at which the use of crystal radiator can be considered reasonable.

The paper is the continuation of the earlier performed the experimental investigations of the spectral and the spectral - angular distributions of the gamma radiation of the electrons with energy about 1000 MeV at their motion along the single crystals axes of the silicon and tungsten in the wide range of the crystals thickness [29-34].

In the paper the experimental gamma radiation spectra of the 1200 MeV electrons in the single crystals of the beryllium 1.2 mm thick, silicon 1.5 mm and 15.0 mm thick and tungsten 1.18 mm thick are presented. About the use prospects of the beryllium crystal as the radiator it was indicated also in paper [19].

The gamma radiation spectra of the 1200 MeV electrons in various registration angles (sr) are presented for the electrons motion along the crystallographic axes: <100>, <110> and <111> of the silicon single crystals 1.5 mm thick.

The gamma radiation spectra of the electrons with energy of 300, 600 and 1200 MeV in the silicon single crystals 15.0 mm thick and tungsten 3 mm thick in the solid angle $0.14\times10^{-6}$ sr are presented.

## 2. SPECTRA AND THE SPECTRAL - ANGULAR DISTRIBUTIONS OF THE ELECTRONS GAMMA RADIATION IN THE BERYLLIUM, SILICON AND TUNGSTEN SINGLE CRYSTALS

In the experiment the gamma radiation spectra of the 1200 MeV electrons in the beryllium single crystal and the spectral - angular distributions in thick single crystals of the silicon and tungsten in the range of the gamma quanta radiation angles $\theta_\gamma = 0...1.28$ mrad were measured.

Under the spectral-angular distributions are understood the electrons radiation spectra wich measured at different radiation angle $\theta_\gamma$ relatively the initial electrons direction (0 rad).

Data are obtained with use of the technique allowing to measure of the gamma quanta spectra without the distortions because of the multiple photons production by one electron in thick single crystals [32, 35].

In Fig. 1 the intensity spectra $\Delta N \cdot \omega/\Delta \omega$ of the gamma radiation of 1200 MeV electrons are given in the Be single crystal 1.2 mm thick in the solid angle $\Delta\Omega = 0.14\times10^{-6}$ sr at their passing along the axis <0001> and plane (01$\bar{1}$0). Here $\Delta N$ - number of the quanta on one incident electron in the channel of the analyzer with width on energy $\Delta\omega$, $\omega$ - average energy of the relevant channel.

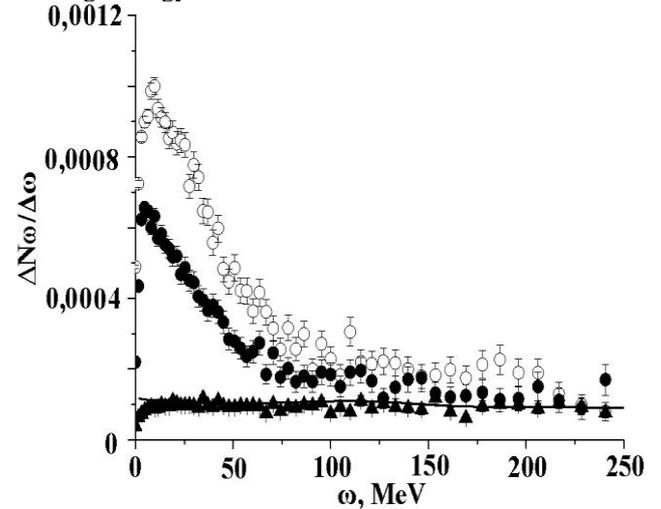

*Fig. 1. The gamma radiation spectra of 1200 MeV electrons to the solid angle $\Delta\Omega=0,14\times10^{-6}$ sr in the Be crystal 1.2 mm thick: ○ - axis <0001>, ● - plane (01$\bar{1}$0), ▲ random crystal. Solid line – calculation on GEANT-4.9.2 for random crystal*

As can be seen from Fig. 1 intensity in the maximum of the gamma radiation spectrum at the axial electron motion approximately by 10 times exceeds intensity for the random oriented crystal, and for the plane motion - approximately by 7 times. The maximum in the intensity spectrum is at the gamma quanta energy $\omega_{max} \sim$ 10 МэВ in the case of the axial electrons motion and $\omega_{max} \sim$ 5 МэВ in case of the plane motion.

Spectral - angular distributions of the gamma radiation was measured for the silicon and tungsten single crystals. Radiation spectra in the solid angles $\Delta\Omega = 0.14\times10^{-6}$, $1.28\times10^{-6}$, $3.55\times10^{-6}$ и $6.97\times10^{-6}$ sr was obtained.

Gamma radiation spectra of the 1200 MeV electrons in the silicon crystals 1.5 mm and 15.0 mm thick and in the tungsten crystal 1.18 mm thick in the radiation solid angles $0.14\cdot10^{-6}$ and $6.97\cdot10^{-6}$ sr are shown in Fig. 2 and 3, respectively.

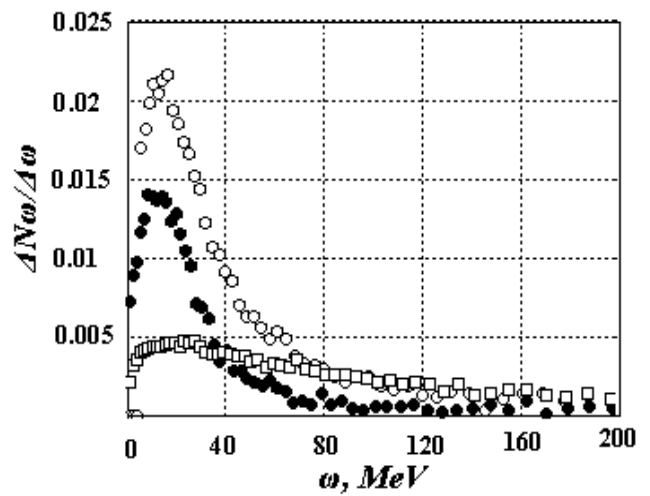

*Fig. 2. The gamma radiation spectra of 1200 MeV electrons to the solid angle $\Delta\Omega=0.14\times10^{-6}$ sr in the Si crystal along axis <111>: ● - 1.5 mm thick: ○ -15.0 mm thick, □ –in W crystal 1.18 mm thick along axis <100>*

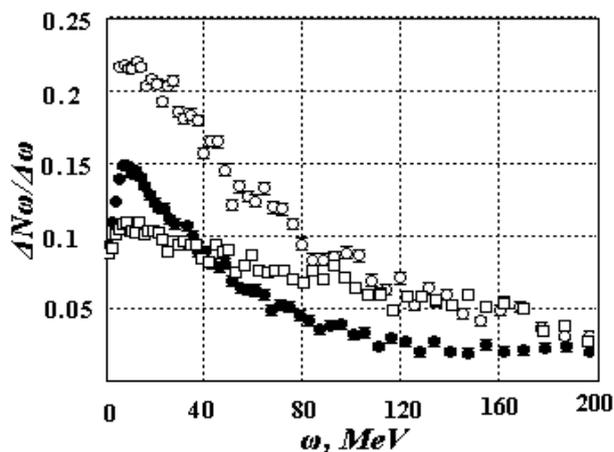

*Fig. 3. The gamma radiation spectra of 1200 MeV electrons to the solid angle ΔΩ=6.97×10⁻⁶ sr in the Si crystal along axis <111>:● - 1.5 mm thick: ○ -15.0 mm thick, □ –in W crystal 1.18 mm thick along axis <100>*

As can be seen from Fig. 2 and 3 the gamma radiation yield for the silicon crystals in the range of the gamma-quanta energy 10...30 MeV is larger, than for the tungsten crystal. Higher energy in the maximum of the gamma radiation intensity spectra in the tungsten crystal in comparison with the silicon crystal makes it more preferable to the positrons generation [16].

From Fig. 2 it is visible that upon the transition from the crystal thickness 1.5 mm to the thickness 15.0 mm at the thickness increase by 10 times the gamma quanta yield increases only approximately by 1.5 times while the ionization losses of electrons increase in proportion to the crystal thickness. This circumstance needs to consider at the choice of the crystal thickness.

## 3. SPECTRAL - ANGULAR DISTRIBUTIONS OF THE ELECTRONS GAMMA RADIATION FOR THREE MAIN AXES OF THE SILICON CRYSTAL

Practically all experimental data on the investigation of the electromagnetic electrons radiation in the silicon single crystal are obtained for the case of the electron beam passing along the axis <111>. However, from the point of view of obtaining the maximum yield of gamma radiation, this axis may not be optimal.

Therefore, spectral-angular distributions of the gamma radiation of the 1200 MeV electrons on three silicon crystal targets 1.5 mm thick were measured. Si crystal targets were cut out from the central part of the silicon ingot so that the directions of the axes <100>, <110> and <111> were perpendicular the geometrical plane of the targets. Spectral-angular distributions of gamma radiation were measured for the angles radiation relatively of the initial electrons beam direction θ$_\gamma$=0, 0.42, 0.85 and 1.28 mrad. Spectral-angular distributions of the gamma radiation from the <111> silicon crystal 1.5 mm thick for the 1200 MeV electrons also were measured earlier and given in [31,32].

From these distributions the gamma radiation spectra of the electrons were obtained in the solid angles of the registration ΔΩ = 0.14·10⁻⁶, 1.28·10⁻⁶, 3.55·10⁻⁶ and 6.97·10⁻⁶ sr at their passing along three main axes <100>, <110> and <111>. Results is shown on Fig. 4(a-d).

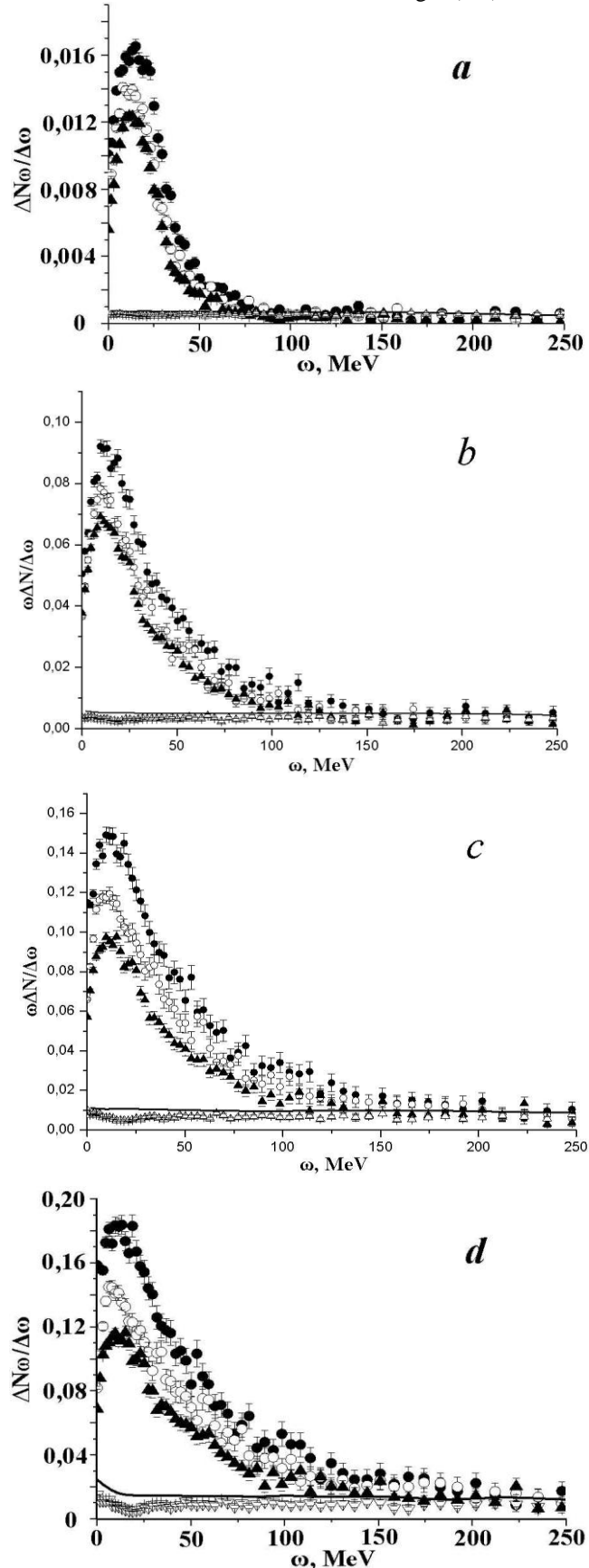

*Fig. 4. The gamma radiation spectra of 1200 MeV electrons in Si crystals 1.5 mm thick: ● - axis <110>, ○ - <111>, ▲ - <100> to the solid angle: a - ΔΩ=0.14·10⁻⁶ sr, b – 1.28·10⁻⁶ sr, c – 3.55·10⁻⁶ sr, d – 6.97·10⁻⁶ sr; Δ – random crystal*

The solid line – calculation for the random crystal on the GEANT-4.9.2 program.

From Fig. 4 can be seen that the gamma radiation spectra of the electrons for different axes of the silicon crystal 1.5 mm thick significantly differ for all the radiation solid angles. The maximal gamma radiation intensity takes place on the electrons passing along the silicon crystal axis <110>. The gamma radiation intensity in this case exceeds, for example, the radiation intensity for the axis <100> approximately on 24% for the radiation solid angle $\Delta\Omega=0.14\cdot10^{-6}$ sr and on 38% for $\Delta\Omega=6.97\cdot10^{-6}$ sr. With decrease of the radiation solid angle is observed also some improvement of the radiation monochromaticity in the range of the radiation intensity maximum.

Measurement of the absolute values of the spectral - angular distributions of the electrons gamma radiation in thick crystals opened the possibility of quantitative comparison of experimental results with the predictions of the theoretical models and hypotheses verification which are the basis of these models.

Calculations were performed taking into account the angular distribution evolution of the electron beam in the crystal, which is bound with the multiple scattering of the particles on the lattice atoms. For the particles passing at the angle $\psi$ to the crystallographic axis exceeding the critical angle of the channeling $\psi_c$ the calculation was performed on the formulas of the modified theory of the coherent radiation; at $\psi<\psi_c$; - the electron trajectory curvature in the field of the crystal atoms raw was considered [31, 33-38]. The contribution of the channeled particles to the gamma radiation was not considered.

The results of the measuring and calculations of the quantity $(\Delta N\times\omega/\Delta\omega/\Delta\Omega)\times10^{-4}$ are presented in Table 1 for three various axes ($\omega$=10 MeV, radiation angle $\theta_\gamma$ =0).

*Table 1. Results of measuring and calculation of the value $(\Delta N\times\omega/\Delta\omega/\Delta\Omega)\times10^{-4}$ for three axes of the Si crystal on $\omega$=10 MeV and angle radiation $\theta_\gamma$=0 for 1200 MeV electrons*

| Axis | Experiment | Calculation |
|---|---|---|
| <111> | 9.2 | 9.5 |
| <110> | 10.3 | 10.7 |
| <100> | 8.3 | 8.7 |

The satisfactory quantitative agreement of the results of the calculation and experiment demonstrates correctness of the assumptions made for the calculations simplification. It means that the regularities of the spectral - angular distributions of the electrons gamma radiation in a thick crystal which were observed in the experiment are caused by the coherent radiation features of the abovebarrier electrons in the field of the crystal atoms raw.

Thus, we can see that in the considered area of the particles energy and the crystals thickness the abovebarrier particles give the main contribution to the angular distributions formation of the gamma radiation. The reason of it, apparently, consists that in the considered energy range the electron motion in the field of the atoms chain in the channeling conditions is extremely unstable. Therefore, there is the fast particles dechanneling, i.e. their transition from subbarrier to abovebarrier states [39].

## 4. MINIMUM ELECTRONS ENERGY FOR THE GAMMA QUANTA SOURCE ON THE CRYSTAL TARGET BASIS

An attempt was made to identify of minimum energy at which application of the crystal radiator as the gamma quanta source for the radiation technologies can be considered still reasonable. The measurements of the gamma radiation spectra of the electrons with energy 300, 600 and 1200 MeV in the silicon single crystals 15.0 mm thick and the tungsten 3.0 mm thick were performed. Radiation Solid angle $\Delta\Omega = 0.14\cdot10^{-6}$ sr. Results are given in Fig. 5 and 6.

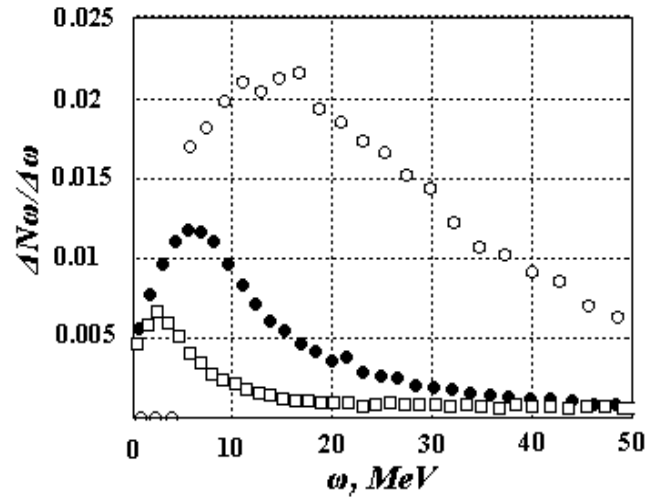

*Fig. 5. The gamma radiation spectra of electrons in Si crystal 15.0 mm thick along axis <111> to solid angle $\Delta\Omega=0.14\cdot10^{-6}$ sr: ○ – E=1200 MeV; ● – E=600 MeV; □– E=300 MeV*

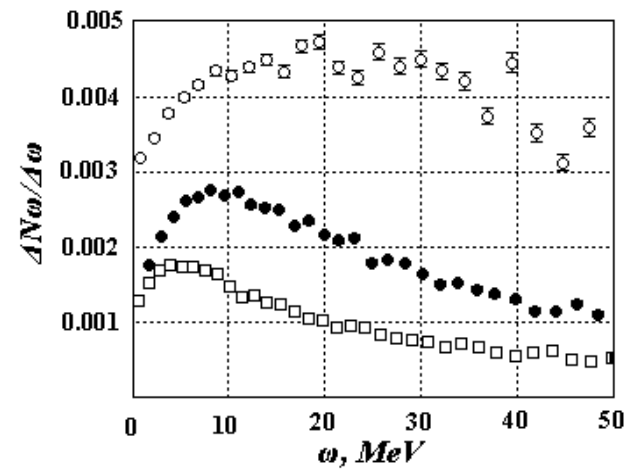

*Fig. 6. The gamma radiation spectra of electrons in W crystal 3.0 mm thick along axis <100> to solid angle $\Delta\Omega=0.14\cdot10^{-6}$ sr: ○ – E=1200 MeV; ● – E=600 MeV; □– E=300 MeV*

From Fig. 5 and 6 it is seen that at the decrease of the electrons energy the maximum of the gamma radiation intensity is shifted in the range of the less energy and the application of the silicon crystal at the electrons energy lower than 600 MeV it is already inexpedient.

However, in the case of the tungsten crystal it is seen that else at the electrons energy 300 MeV the application of the tungsten crystal as the radiator can be still efficient.

## 5. INFLUENCE OF THE OUTPUT FOIL OF THE ACCELERATOR VACUUM CHAMBER

One of the principle opportunities to increase the electrons current from the crystal target, and, respectively, the maximum achievable gamma radiation intensity is the forced crystal cooling. For this, it is necessary to install a goniometer with a crystal in the air outside the vacuum chamber of the accelerator.

To check the effect of the output foil of the vacuum chamber on the radiation parameters in front of the crystals, an aluminum foil 0.165 mm thick was installed. This simulated the output foil of the accelerator vacuum chamber. In this geometry gamma-ray spectra were measured. On Fig. 7 the gamma radiation spectra of the 1200 MeV electrons are shown for the axis <111> in the silicon crystals 3.0 mm thick (Fig. 7a) and 1.5 mm thick (Fig. 7b) before (light circles) and after (dark circles) the installation of aluminum foil in front of the crystal.

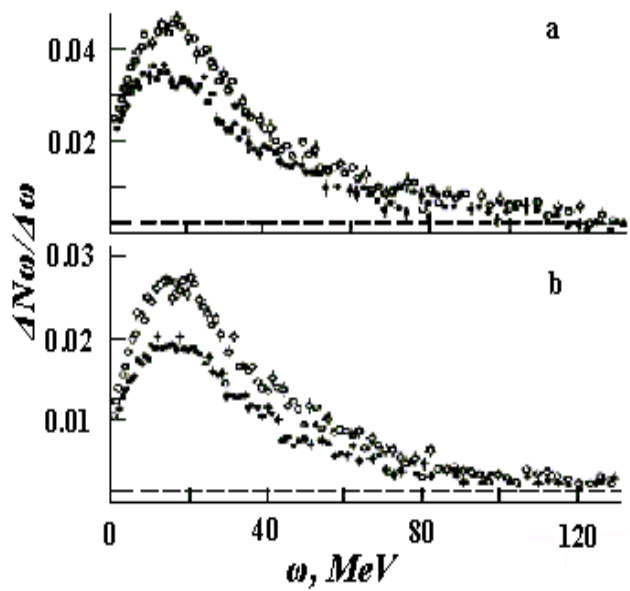

*Fig. 7. The gamma radiation spectra of 1200 MeV electrons passing along axis <111> in Si crystals 3.0 mm thick (Fig. 7a) and 1.5 mm thick (Fig. 7b), before (light circles) and after (dark circles) the installation of aluminum foil 0.165 mm thick in front of the crystals. Dush line – random oriented Si crystal*

From Fig. 7 it is seen that the installation of the goniometer with the crystal out the accelerator vacuum chamber leads to the essential decrease of the gamma radiation intensity: about 21% for the crystal 3.0 mm and about 28% for the crystal 1.5 mm thick. However, this decrease of the intensity can be compensated by the increase of the electrons current due to the possibility of the forced cooling of the crystal target and simplicity of the goniometer maintenance.

## CONCLUSION

Gamma radiation spectra of the 1200 MeV electrons for various radiation angles from silicon crystals 1.5 mm and 15.0 mm thick and tungsten crystal 1.18 mm thick with orientation of the axes <111> and <100> are measured.

It is shown that for the silicon crystals were produced larger gamma radiation yield in the energy range of gamma quanta 10 … 30 MeV than for the tungsten crystal. Higher energy in the maximum of the gamma radiation intensity in W crystal in the comparison with Si crystal does W more preferable for the positrons generation.

Spectral-angular distributions of the gamma radiation of the 1200 MeV electrons on three silicon crystal targets 1.5 mm thick with the orientation <100>, <110> and <111> axes were measured. Spectral-angular distributions of gamma radiation were measured for solid angles $\Delta\Omega = 0.14\cdot10^{-6}$, $1.28\cdot10^{-6}$, $3.14\cdot10^{-6}$ and $6.97\cdot10^{-6}$ sr. These corresponded with angles relatively of the initial electrons beam direction $\theta_\gamma$=0, 0.42, 0.85 and 1.28 mrad.

It was established experimentally that ehe maximal gamma radiation intensity takes place on the electrons passing along the silicon crystal axis <110>. The gamma radiation intensity in this case exceeds, for example, the radiation intensity for the axis <100> approximately on 24% for the radiation solid angle $\Delta\Omega=0.14\cdot10^{-6}$ sr and on 38% for $\Delta\Omega=6.97\cdot10^{-6}$ sr. With decrease of the radiation solid angle is observed also some improvement of the radiation monochromaticity in the range of the radiation intensity maximum.

The measurements of the gamma radiation spectra of the electrons with energy 300, 600 and 1200 MeV in the silicon single crystals 15.0 mm thick and the tungsten 3.0 mm thick were performed. It is shown that still at the electrons energy 300 MeV the application of the tungsten crystal the radiator can be efficient for the obtaining of the radiation beams for radiation technologies.

The above results complement the previously obtained experimental and theoretical data and can be used to create a source of intense gamma-ray beams for radiation technologies on the basis of the crystal targets and linac with electrons energy about 1000 MeV.